\documentstyle[aps, floats]{revtex}
\begin{document}
\narrowtext
\twocolumn

{\bf Comment on
 ``Two Phase Transitions in the Fully frustrated XY Model"}
\vspace{0.1cm}

The fully frustrated two-dimensional XY model (FFXYM) has been the source
of many recent studies, and controversies 
\cite{olsson,lgk91gn93lee94,lee,grsjjv92,gkncm95,physica}. 
The reason for this interest is that this model appears to have 
rather unconventional  critical properties. 
Furthermore, this model has a clear experimental representation in
Josephson junction arrays in magnetic fields. The model has U(1) and
Z(2) symmetries, associated to its phase and chiral, or Ising, degrees of 
freedom, respectively. These symmetries are intertwined in
purely thermodynamic quantities, like the helicity modulus, $\Upsilon$.
The possibility of defining a new universality class to 
describe the properties of the model appears to depend on having 
or not both symmetries breaking at the same temperature. Answering 
this question conclusively has turned out to be very hard since 
both transitions seem to occur either very close or may be 
on top of each other.

In a recent Letter \cite{olsson} Olsson has argued that (i) both 
transitions take place at different temperatures, (ii) 
that the jump in $\Upsilon_{FFXYM}$ is of the standard 
unfrustrated 2-D XY Berezinskii-Kosterlitz-Thouless 
(BKT) type and (iii) that the chiral exponent $\nu_{Z(2)}=1$, as 
in the standard 2-D Ising model,  contradicting a large 
number of previous papers that obtained
$\nu_{Z(2)}<1$ \cite{lgk91gn93lee94,grsjjv92,gkncm95,physica}. 
Here we argue that the analysis
given to reach these conclusions has flaws and the conclusions 
are not consistent with the evidence presented. Furthermore, we
present new data of a more extensive MC calculation that gives
$\nu_{Z(2)}<1$. 

(i-ii) The conclusion of having two different transition 
temperatures is based in part on the {\underline {assumption}} that 
$\Upsilon_{FFXYM}(T_c)\equiv \Upsilon (T_{BKT})$. 
This translates into setting the exponent $\eta=\frac{1}{4}$ exactly.
%It is hard to see how could this be the case when both models have
%such different types of elementary excitations, one with long range
%order and the other with only a topological one.  
A very important element missing in the analysis of \cite{olsson}
is that there is no explicit mention of any type of error analysis 
carried out in the calculations. It is well known that this type 
of analysis is crucial in reliably determining exponents 
in XY-like models via MC calculations \cite{gupta}.  
Then the claim that the two critical temperatures are different an equal to
$T_{BKT}=0.446$ and $T_{Z(2)}=0.452$, which differ by about 
$3\%$,  without quoting any error bars appears to be overly 
optimistic. A similar and analysis, including errors, was done in \cite{lee},
where they instead find a {\it non-universal} jump of $0.22(2)$.
In contrast to \cite{olsson,lee} we have carried out an extensive 
explicit calculation of the gauge  invariant zero momentum U(1) 
correlation function and  we found {\it instead} 
$\eta\approx\frac{1}{5}$ \cite{grsjjv92}. Of significant importance 
is that this result {\underline {has been}} successfully compared with 
experimental results in JJA \cite{physica}.

(iii)  Olsson calculated the chiral correlation function and then  
its coherence length $\xi_{Z(2)}$, from which he got 
$\nu_{Z(2)}=1$. One important drawback of this calculation is that the data 
analyzed was taken about $15\%$ away from  $T_{Z(2)}=0.452$.
In \cite{grsjjv92} we concentrated in the calculation of
the $U(1)$ correlation functions and less so in the chiral one. We have now
extended our analysis of the zero momentum chiral correlation function,
and the  $\xi_{Z(2)}(T)$   data is shown in the table.
\begin{table}[hd]
\caption {Results for $\xi_{Z(2)}$ obtained from $g_{z(2)}(r)$ at
different temperatures and lattice sizes.}
\begin{center}
\begin{tabular}{ c c c c c c c }
\hline
$T$   & $L$ & $\xi_{Z(2)}$   & NMCS & $L$ & $\xi_{Z(2)}$  & NMCS   \\
      &     &                &  K   &     &               &  K     \\
\hline
0.70  &  24 & 0.8217 (671)   & 160  &  32 &  0.9312 (92 )   &  160 \\
0.60  &  24 & 1.3341 (68 )   & 160  &  32 &  1.5520 (530)   &  160 \\  
0.575 &  24 & 1.4318 (73 )   & 160  &  32 &  1.7579 (62 )   &  160 \\
0.55  &  24 & 1.5386 (43 )   & 160  &  32 &  2.1278 (26 )   &  160 \\   
0.525 &  24 & 2.3948 (37 )   & 175  &  32 &  2.9035 (101)   &  175 \\ 
0.50  &  48 & 3.8496 (37 )   & 175  &  64 &  4.4840 (38 )   &  150 \\ 
0.49  &  48 & 4.3700 (36 )   & 225  &  64 &  5.9480 (37 )   &  225 \\ 
0.48  &  48 & 5.3916 (20 )   & 225  &  64 &  7.2132 (432)   &  225 \\ 
0.47  &  48 & 6.4206 (82 )   & 225  &  64 &  9.8993 (586)   &  225 \\ 
0.46  &  -- & ------------   & ---  & 128 & 20.2928 (464)   &  175 \\  
\hline
\end{tabular}
\end{center}
\end{table}
We have carefully analyzed the data,
together with an error analysis, as done in \cite{grsjjv92,gupta}, in an 
extended temperature range, and we obtained $\nu_{Z(2)}=0.898(3)$ and 
$T_{Z(2)}=0.4511(10)$. The critical exponent is in good agreement with 
previous calculations while the critical temperature is close to the one
obtained in \cite{olsson}. Furthermore, recent experiments in 
superconducting wire networks \cite{paul}, where the order parameter 
fluctuations are important, have also found that the 
chiral exponent $\nu_{Z(2)}\neq 1$. 
\vskip 0.1 cm
This work has been partially supported by
grant NSF-DMR-95-21845 (JJV), and by DGAPA-UNAM grants IN-103294 and
IN-100595 (GRS). The calculations were mostly done at DGSCA-UNAM.
\vskip 0.2cm
Jorge V. Jos\'e$^{1}$ and G. Ram\'\i rez-Santiago$^{2}$
\vskip 0.1cm
$^1$ Department of Physics and Center for Interdisciplinary 
Research on Complex Systems, Northeastern University, Boston MA, 02115.
\vskip 0.001cm
$^2$ Instituto de F\'\i sica, UNAM, P.O. Box 20-364, 01000 M\'exico D.F.

\end{document}